\documentclass[12pt,a4paper]{article}

\RequirePackage[l2tabu, orthodox]{nag}
\usepackage{diagbox}
\usepackage{silence}
\usepackage{multirow}
\ActivateWarningFilters[pdftoc]

\usepackage[T1]{fontenc}
\usepackage[utf8]{inputenc}

\usepackage{subcaption}

\newcommand \versionno{\jobname.tex}

\usepackage{array,longtable}
\usepackage{dsfont}

\usepackage[margin=20mm]{geometry}

\usepackage{pgf}
\usepackage{tikz}
\usepackage{braket}
\usepackage{xcolor}
\usepackage{yfonts}

\usepackage{amsmath,amssymb,amsfonts,amscd,latexsym,amsthm}

\usepackage[colorlinks=true, linktoc=all, linkcolor=blue, citecolor=red, urlcolor=blue, backref=page]{hyperref}
\usepackage{dsfont}

\usepackage{pgf}
\usepackage{tikz}
\usetikzlibrary{calc}

\usepackage{tikz}
\usetikzlibrary{hobby}
\usetikzlibrary {decorations.pathmorphing}

\graphicspath{ {./plots/} }

\usepackage{etoolbox}
\makeatletter
\patchcmd{\BR@backref}{\newblock}{\newblock(}{}{}
\patchcmd{\BR@backref}{\par}{)\par}{}{}
\makeatother

\numberwithin{equation}{section}
\newtheoremstyle{break}{9pt}{9pt}{\itshape}{}{\bfseries}{}{\newline}{}
\theoremstyle{break}

\begin{document}

\begin{flushleft}
{\bfseries\sffamily\Large 
Boundary operators in the Brownian loop soup
\vspace{1.5cm}
\\
\hrule height .6mm
}
\vspace{1.5cm}

{\bfseries\sffamily 
Federico Camia$^{1,2}$
Rongvoram Nivesvivat$^{1}$
}
\vspace{4mm}

{\textit{\noindent
$^1$ 
New York University Abu Dhabi, Abu Dhabi, United Arab Emirates
\\
$^2$ 
New York University, New York, United Sates of America
}}
\vspace{4mm}

{\textit{E-mail:} \texttt{
federico.camia@nyu.edu,
rongvoramnivesvivat@gmail.com
}}

\end{flushleft}
\vspace{7mm}

{\noindent\textsc{Abstract:}
We obtain infinitely many boundary operators in the Brownian loop soup in the subcritical phase by analyzing the conformal block expansion of the two-point function that computes the probability of having two marked points on the upper half-plane being separated by Brownian loops. The resulting boundary operators are primary operators in a 2D CFT with central charge $c\leq1$ and have conformal dimensions that are non-negative integers. By comparing the above-mentioned conformal block expansion with probabilities in the Brownian loop soup, we provide a physical interpretation of the boundary operators of even dimensions as operators that insert multiple outer boundaries of Brownian loops at points on the real axis.
\clearpage


\hrule 
\tableofcontents
\vspace{5mm}
\hrule
\vspace{5mm}

\hypersetup{linkcolor=blue}

\section{Introduction}

In \cite{LW03}, Lawler and Werner proposed a conformally-invariant statistical model, defined as a Poisson ensemble of Brownian loops on the plane, and named it the Brownian loop soup. In this model, Brownian loops are defined as Brownian paths starting and ending at the same point. Since its introduction, the Brownian loop soup has been much studied due to its connections with other interesting objects in probability theory and statistical mechanics. For instance, the outer boundary of a Brownian loop is distributed like the scaling limit of the outer boundary of a critical percolation cluster, which is also believed to coincide with the conjectured scaling limit of the self-avoiding walk and with the $O(n\rightarrow 0)$ loop model of \cite{Nienhuis82}. Moreover, the collection of outer boundaries of clusters of Brownian loops (where any two intersecting Brownian loops belong to the same cluster) is a Conformal Loop Ensemble (CLE) \cite{cn04,she06,sw11}, a conformally invariant collection of random planar loops that are locally distributed like SLE curves. The latter, denoted by SLE$_\kappa$, form a one-parameter family of planar curves that obey the so-called Schramm-Loewner Evolution \cite{schram99} and are the (conjectural) scaling limits of interfaces of several models of statistical mechanics, including percolation \cite{smirnov01,cn07} and the Ising model \cite{CDCHKS14}.

More recently, it was found in a series of papers \cite{cgk16, cfgk20, cfgk21, cfgk22} that there is also a correspondence between some observables of the Brownian loop soup and conformally-covariant correlation functions. This suggests the existence of a conformally-invariant quantum field theory in two dimensions, a 2D conformal field theory (CFT), that describes aspects of the Brownian loop soup. However, it remains unclear whether it is always possible to recast any Brownian loop soup observable in terms of correlation functions of a CFT. The goal of this note is to further explore the connection between boundary local operators in 2D CFT and the Brownian loop soup on the upper half-plane.

\subsection{Brief review of the model}
The Brownian loop soup in a domain $D$ (later, we will restrict our attention to the upper half-plane, $D=\mathbb{H}$) is a Poissonian ensemble of loops with intensity $\lambda>0$, where $\lambda$ can be equivalently interpreted as the fugacity of each Brownian loop in a grand canonical ensemble of loops whose partition function can be written as
\begin{equation}
Z_{\text{BLS}}
=\sum_{k=0}^\infty \frac{\lambda^k}{k!}(\mu^{\text{loop}}_{D})^k
\ ,
\label{BLS}
\end{equation}
where $\mu^{\text{loop}}_{D}$, called the Brownian loop measure in $D$, is the intensity measure used in~\cite{LW03} in the definition of the Brownian loop soup as a Poisson process of planar loops.
The Brownian loop measure $\mu^{\text{loop}}_{D}$ is related to the two-dimensional Brownian bridge measure, the natural measure on Brownian paths starting and ending at the same point. 
Up to a multiplicative constant, it is essentially the  only nontrivial measure on loops (closed planar curves) that is invariant under all conformal transformations, both global and local (for a precise definition of $\mu^{\text{loop}}_{D}$ and its properties, see~\cite{w06}).

The model described by the partition function in \eqref{BLS} has two physical phases, depending on $\lambda$: the supercritical phase for $\lambda > \frac12$ and the subcritical phase for $\lambda \leq \frac12$. In the supercritical phase, there is only 1 cluster of Brownian loops (that is to say, any two loops are connected by a chain of loops that intersect each other) and the complement of the union of all loops is a totally disconnected set (i.e., it does not contain sets larger than single points---see \cite{bc10}). In the subcritical phase, there are infinitely many clusters of Brownian loops, whose outer boundaries are locally distributed like SLE$_\kappa$ curves with $\kappa \in (\frac83, 4]$,  where $\kappa$ depends of $\lambda$ according to the relation \cite{w03}
\begin{align}
\lambda &= \frac{(6 - \kappa)(3\kappa - 8)}{4\kappa}
\ .
\end{align}

We are interested in the subcritical phase of the Brownian loop soup, which is expected to be described by a 2D CFT with central charge $c \leq 1$, given by
\begin{align}
c &= 2\lambda
 \ .
 \label{cl}
\end{align}

It is also worth mentioning that, in the limit $\lambda \rightarrow 0$, the leading term of the partition function \eqref{BLS} contains only contributions with a single Brownian loop, whose outer boundary, as mentioned earlier, is conjectured to coincide with the self-avoiding loop of the $O(n\rightarrow 0)$ loop model of \cite{Nienhuis82}. The coincidence between the two models in this limit was exploited in \cite{cfgk20} to express several correlation functions of the Brownian loop soup in terms of correlation functions of the so-called twist operator in the $O(n \rightarrow 0)$ loop models, first computed in \cite{cs09}.

\subsubsection{Previous CFT results for the Brownian loop soup}
A 2D CFT can be defined as a collection of local operators that transform according to the two-dimensional conformal algebra, also known as Virasoro algebra, and also satisfy consistency conditions such as the conformal Ward identities and the associativity expressed by the operator product expansion (OPE) \cite{bpz84}. One of the main goals of CFT, with important implications in statistical mechanics, is a complete classification of all universality classes in terms of representations of the infinite-dimensional Virasoro algebra. 

Thus, the first step to find a CFT describing observables in the Brownian loop soup is to identify an appropriate collection of local operators. 
In this direction, in~\cite{cgk16, cfgk21}, the authors proposed two families of bulk local operators, which we review below.
\begin{itemize}
\item[1.]   {\em Layering operators} $O_{\beta}(z)$: primary operators with left- and right-conformal dimensions $( \Delta_{{\beta}},  \Delta_{{\beta}})$, where
\begin{equation}
\Delta_{{\beta}}= \frac{\lambda}{10}(1 - \cos\beta)
\ .
\end{equation}
In terms of Brownian loops, $O_{\beta}(z) = \sum_{\gamma}\sigma_{\gamma}$, where the sum is over all Brownian loops $\gamma$ that wind around $z$ and $\sigma_{\gamma}=\pm 1$ is a random sign associated to each loop $\gamma$. We will discuss this type of operator in detail in Section \ref{sec:lay}. 


\item[2.]  {\em Charged-edge operators} $\mathcal{E}^{(k)}_\beta(z)$: primary operators with left- and right-conformal dimensions $(\Delta_\beta + \frac k3, \Delta_\beta + \frac k3)$ with $k=1,2,3\ldots$ .  For $\beta=0$, the operator $\mathcal{E}^{(k)}_0(z)$ essentially inserts $k$ outer boundaries of Brownian loops through $z$, and we simply call $\mathcal{E}^{(k)}(z) \equiv \mathcal{E}^{(k)}_0(z)$ an edge operator. In the case $\beta \neq 0$, the operator gets a ``twist'' by assigning a phase to each loop that winds around $z$.

\end{itemize}
The definitions of these operators are ambiguous due to both infrared and ultraviolet divergences, but can be made precise with the introduction of cutoffs, which can then be removed within correlation functions \cite{cgk16, cfgk21}.

Previous results on local operators in the Brownian loop soup concern mostly bulk operators, with the exception of the results of~\cite{cfgk22} on the boundary edge operator. The goal of this note is to further explore the spectrum of boundary operators and their interpretation in terms of Brownian loops. 

\subsection{Main results and plan}
\begin{itemize}
\item In Section \ref{sec:lay}, we decompose the two-point function of
the layering operator $O_{\beta}(z)$ on the upper half-plane into conformal blocks. Let us summarize the main results of this section.
\begin{itemize}
\item[1.] 
We write the OPE between $O_{\beta_1}(z_1)$ and $O_{\beta_2}(z_2)$,
 \begin{multline}
O_{\beta_1}(z_1)  O_{\beta_2}(z_2) 
\\
\overset{z_1 \rightarrow z_2} = 
\sum_{k=0}^\infty
|z_1 - z_2|^{2\Delta_{12} + \frac{2 k }{3} -2\Delta_{\beta_1} -2\Delta_{\beta_2} }
\hat{C}_{O_{\beta_1} O_{\beta_2} \mathcal{E}^{(k)}_{\beta_1 + \beta_2}}
\mathcal{E}^{(k)}_{\beta_1 + \beta_2}(z_2)
+ \ldots
\label{OPE1}
\ ,
 \end{multline}
 where $\Delta_{12}$ is defined in \eqref{12}, and $\hat{C}_{O_{\beta_1} O_{\beta_2} \mathcal{E}^{(k)}_{\beta_1 + \beta_2}}$ are the OPE coefficients, which can be determined on a case-by-case basis from the conformal block expansion. 
 The main difference between the upper half-plane OPE in \eqref{OPE1} and its full-plane counterpart in \cite{cfgk20} is the absence of primary operators with non-zero spin in \eqref{OPE1}, required by the conformal invariance on the upper half-plane.

\item[2.] 
We write the bulk-boundary OPE of $O_{\beta}(z)$,
\begin{multline}
O_{\beta}(z) 
\overset{z \rightarrow \bar z}= 
|z - \bar z|^{ -2\Delta_\beta }
\Bigg(
1+
|z - \bar z|^{2}
\left\{
\frac{\Delta_\beta}{\lambda}
T(x) + b^{(2)}_\beta Y(x)
\right\}
\\
+ 
\sum_{k=3}^\infty
|z - \bar z|^{ k}
b^{(k)}_\beta
e^{(k)}(x)
+ \ldots
\Bigg)
\ ,
\label{OPEbd}
\end{multline}
where $x = \frac{z + \bar z}{2}$ and $T(x)$ is the boundary stress-energy tensor. The coefficient $b^{(k)}_\beta$ is the bulk-boundary OPE coefficient between $O_{\beta}(z) $ and the primary boundary operators. In practice, we can compute $b^{(k)}_\beta$ on a case-by-case basis from the conformal block expansion. 

\item[3.]
For $\beta=\pi n, n \in \mathbb{Z}$, some of the OPE coefficients appearing in~\eqref{OPEbd} vanish:
\begin{equation}
b^{(k)}_{\pi n}
= 0 
\quad\text{for}\quad n\in\mathbb{Z}, k = 2 \;\;\text{and}\;\; 3, 5, 7, \ldots
\ .
\label{conb}
\end{equation}
When $\beta=2\pi n, n\in\mathbb{Z}$, the vanishing of the OPE coefficients above is trivial and $O_{\beta}(z)$ is the identity operator.
For $\beta=(2n+1)\pi, n\in\mathbb{Z}$, the vanishing of the OPE coefficients in \eqref{conb} is not trivial and is deduced from~\eqref{cond}, which is itself conjectured based on~\eqref{D1-7}.

\item[4.]
For generic $\beta$, the OPE \eqref{OPEbd} has two operators of dimension $2$: $T(x)$ and $Y(x)$. We do not know the physical meaning of $Y(x)$ in terms of Brownian loops, but for $\beta=\pi n, n \in \mathbb{Z}$, $Y(x)$ is not present in the OPE~\eqref{OPEbd}, due to~\eqref{conb}.

\item[5.]
From the OPE \eqref{OPEbd}, we propose new boundary primary operators, $Y(x)$ and $e^{(k)}(x)$, whose dimensions are
\begin{equation}
\Delta^{(Y)}
=2
\quad\text{and}\quad
\Delta_k^{(e)} = k 
\quad\text{for}\quad
k = 3, 4, 5 \ldots
\ .
 \label{loc}
\end{equation}

\end{itemize}

\item In Section \ref{sec:prob}, we write the conformal block expansion of $\langle O_{\beta}(z_1)  O_{-\beta}(z_2) \rangle_{\mathbb{H}}$ 
in terms of the Brownian loop measure $\mu_{\mathbb{H}}^{\text{loop}}$. This provides an interpretation of $e^{(2k)}(x)$ as an operator that inserts $k$ Brownian loops that touch the real axis at $x$.

\item
In Section \ref{sec:newE}, we discuss the statistical interpretation of the edge operators $\mathcal{E}^{(k)}_\beta(z)$.
The OPE \eqref{OPE1} suggests that the one-point functions $\langle \mathcal{E}^{(k)}_0(z)\rangle$ are non-zero. This is inconsistent with the mathematical definition of $\mathcal{E}^{(k)}_0(z)$ in~\cite{cfgk20}, which gives $\langle\mathcal{E}^{(k)}_0(z)\rangle=0$. We propose a new definition of $\mathcal{E}^{(k)}_0(z)$ on the upper half-plane whose one-point function is not identically zero.
\item
In Section \ref{sec:out}, we discuss the outlook of these results.
\end{itemize}

\section{Two-point functions of the layering operators \label{sec:lay}}
The goal of this section is to study the OPEs of the layering operators ${O}_\beta(z)$. In particular, we aim to write down the list of local boundary operators whose dimensions are listed in~\eqref{loc} by considering the bulk-boundary OPE of $O_\beta (z)$: in the limit $z \rightarrow \bar z$, we expand ${O}_\beta(z)$ as an infinite series of those boundary operators. In practice, this can be done by considering the conformal block expansion of the two-point function of ${O}_\beta(z)$. From \cite{cfgk20}, we have 
\begin{equation}
\langle 
O_{\beta_1}(z_1)O_{\beta_2}(z_2)
\rangle_{\mathbb{H}}
= 
|z_1 - \bar z_2|^{-4\Delta_{\beta_1}}|z_2 - \bar z_2|^{2\Delta_{\beta_1} - 2\Delta_{\beta_2}}G_{\beta_1, \beta_2}(\sigma)
\ , 
\label{2pts}
\end{equation}
where the cross-ratio $\sigma$ is defined by
\begin{equation}
\sigma = \frac{|z_1 - z_2|^2}{|z_1 - \bar z_2|^2}
\label{cross}
\end{equation}
and the function $G_{\beta_1, \beta_2}(\sigma)$ is given by
\begin{multline}
G_{\beta_1, \beta_2}(\sigma)
= \sigma^{\Delta_{12} - \Delta_{\beta_1} - \Delta_{\beta_2}}(1- \sigma)^{-2\Delta_{\beta_1}}
\text{exp}
\left\{
(\Delta_{12} -\Delta_{\beta_1} - \Delta_{\beta_2} )(1-\sigma)
{}_3F_2\left( 1, 1, \frac43; 2, \frac53; 1-\sigma \right)
\right\}
\ ,
\end{multline}
where we have introduced the notation
\begin{subequations}
\begin{align}
\Delta_{ij} &= \frac{\lambda}{10}(1 - \cos(\beta_i + \beta_j))
\ .
\label{12}
\end{align}
\end{subequations}
Furthermore, notice that the two-point function \eqref{2pts} is invariant under $\beta_i \rightarrow \beta_i + 2\pi \mathbb{Z}$.

Two-point functions of bulk operators on the upper half-plane and chiral four-point functions on the full plane $\mathbb{C}$ obey the same conformal Ward identities \cite{cardy84}.
More precisely, if $o_{\beta_i}(z)$ denotes a chiral operator of dimension $\Delta_{\beta_i}$, the correlation functions
\begin{equation}
\langle 
O_{\beta_1}(z_1)O_{\beta_2}(z_2)
\rangle_{\mathbb{H}}
\quad\text{and}\quad
\langle o_{\beta_1}(z_1)o_{\beta_2}(z_2)o_{\beta_2}(\bar z_2)o_{\beta_1}(\bar z_1) \rangle_{\mathbb{C}}
\label{24re}
\end{equation}
obey the same conformal Ward identities.

Next, recall that chiral four-point functions on the full plane always admit a decomposition in terms of conformal blocks~\cite{rib14}. Therefore, we can expand the two-point function in~\eqref{2pts} in conformal blocks~\cite{zam84}. With this, we can write the limit $z_1 \rightarrow z_2$ of the two-point function \eqref{2pts} as an infinite series of the $s$-channel conformal blocks of the four-point function in \eqref{24re}, whereas the limit ${z_2 \rightarrow \bar z_2}$ of \eqref{2pts} can be expanded in the $t$-channel conformal blocks. We now analyze the two limits of \eqref{2pts} in detail.

\subsection{The $s$-channel expansion}
We start with the conformal block expansion corresponding to the limit $z_1 \rightarrow z_2$ of the two-point function \eqref{2pts}, that is to say, we perform the OPE between the two layering operators in \eqref{2pts} as $z_1 \rightarrow z_2$.
In practice, this is equivalent to taking the limit $\sigma \rightarrow 0 $ in \eqref{2pts}. First, we consider the limit $\sigma \rightarrow 0$ of $G_{\beta_1, \beta_2}(\sigma)$. We deduce
\begin{equation}
G_{\beta_1, \beta_2}(\sigma) \overset{\sigma \rightarrow 0}{=} \sigma^{-\Delta_{\beta_1} - \Delta_{\beta_2}} \sum_{k \in\frac13 \mathbb{N}}A_k \sigma^{k}
\ ,
\label{Gs}
\end{equation}
where 
$\mathbb{N}$ denotes the set of non-negative integers. The first few coefficients $A_k$ in the above expansion are
\begin{subequations}
\begin{align}
A_0&=
e^{-\frac{2 \pi  \left(\Delta _{\beta _1}+\Delta _{\beta _2}-\Delta _{12}\right)}{\sqrt{3}}}\ , \\
A_{\frac13} &= 
\left(\Delta_{12}-\Delta _{\beta _1}-\Delta _{\beta _2}\right)
\frac{2^{2/3} \Gamma \left(-\frac{1}{3}\right) \Gamma \left(\frac{5}{6}\right)}{\sqrt{\pi }}A_0
\ , \\
A_{\frac23} &=
 \frac{(A_{\frac 13})^2}{2!}
\ , \\
A_{1} &=
\left(\Delta_{12}+\Delta _{\beta _1}-\Delta _{\beta _2}\right)A_0 +  \frac{(A_{\frac 13})^3}{3!}
\ , \\
A_{\frac 43} &=
\left(\frac16 + \Delta_{12}+\Delta _{\beta _1}-\Delta _{\beta _2}\right)A_1 +  \frac{(A_{\frac 13})^4}{4!}
\ .
\end{align}
\end{subequations}
Therefore, the expansion \eqref{Gs} implies that the two-point function \eqref{2pts} has the following $s$-channel decomposition,
\begin{equation}
\langle 
O_{\beta_1}(z_1)
O_{\beta_2}(z_2)
\rangle_{\mathbb H}
\overset{z_1 \rightarrow z_2}{=}
|z_1 - \bar z_2|^{-4\Delta_{\beta_1}}|z_2 - \bar z_2|^{2\Delta_{\beta_1} - 2\Delta_{\beta_2}}
\sum_{k\in \mathbb{N}}
d_{\Delta_{12} + \frac k3}\mathcal{F}_{
\Delta_{12} + \frac k 3}^{(s)}(\Delta_{\beta_1}, \Delta_{\beta_2}, \Delta_{\beta_2}, \Delta_{\beta_1}| \sigma)
\ ,
\label{sch}
\end{equation}
where the function $\mathcal{F}_\Delta^{(s)}(\Delta_{\beta_1}, \Delta_{\beta_2}, \Delta_{\beta_2}, \Delta_{\beta_1}| \sigma)$ denotes the $s$-channel conformal block of dimension $\Delta$, $\sigma$ is the cross-ratio defined in~\eqref{cross}, and $d_{\Delta_{12} + \frac k3}$ are the structure constants to be determined. The conformal dimensions appearing in the sum on the right-hand side of \eqref{sch} were deduced from the powers of $\sigma$ in \eqref{Gs}. They are the conformal dimensions of bulk operators that appear in the OPE \eqref{OPE1}.

To determine $d_{\Delta_{12} + \frac k3}$ in \eqref{sch}, we simply compare the expansion \eqref{sch} to the two-point function in \eqref{2pts}. To start, we recall from \cite{zam84} that the $s$-channel conformal block can be written as an infinite power series around $\sigma = 0$,
\begin{equation}
\mathcal{F}_\Delta^{(s)}(\Delta_{\beta_1}, \Delta_{\beta_2}, \Delta_{\beta_2}, \Delta_{\beta_1}| \sigma)
=
\sigma^{\Delta - \Delta_{\beta_1} - \Delta_{\beta_2}}
\sum_{k=0}^\infty a_k \sigma^k
\ .
\label{sa}
\end{equation}
There are many ways to compute the coefficients $a_k$ in \eqref{sa}, for instance, $a_k$ can be computed as solutions to the local conformal Ward identities \cite{rib14}. In this note, we choose to use the Zamlodchikov recursion \cite{zam84} of conformal blocks---see Appendix~\ref{ap:Vir}. In principle, this recursion allows us to determine all coefficients $a_k$ recursively.

Comparing equations \eqref{2pts}, \eqref{sch}, \eqref{sa} and \eqref{Gs}, we obtain a system of linear equations with the structure constants $d_{\Delta_{12} + \frac k3}$ as unknowns. 
For non-negative integers $k$, we define the function 
\begin{equation}
h\left(\frac k3\right) = 
\sum_{\substack{\frac{j}{3} + m = \frac{k}{3} 
\\
j,m\in\mathbb{N}
}}d_{\Delta_{12} + \frac{j}{3}} a_m - A_{\frac{k}{3}}
\ .
\end{equation}
Then, the structure constant $d_{\Delta_{12} + \frac k3}$ is a solution to the equations
\begin{equation}
h\left(\frac k3\right)
=
h\left(\frac {k- 1}3 \right) =
\ldots = 
h\left(\frac 13 \right) = h(0) = 0
\ .
\end{equation}
For instance,  we find
\begin{subequations}
\begin{align}
d_{\Delta_{12} } &= A_0
\ , \\
d_{\Delta_{12} + \frac13} &= A_{\frac13}
\ , \\
d_{\Delta_{12} + \frac23} &=  \frac{(A_{\frac 13})^2}{2!}
\ , \\
d_{\Delta_{12} + 1} &=  \frac{(A_{\frac 13})^3}{3!} + \frac{(\Delta_{\beta_2} -\Delta_{\beta_1})(\Delta_{\beta_1} -\Delta_{\beta_2} + \Delta_{12} )}{\Delta_{12}}A_0
\ , \\
d_{\Delta_{12} + \frac43} &=  \frac{(A_{\frac 13})^4}{4!} + \left\{
\Delta_{\beta_1} -\Delta_{\beta_2} + \Delta_{12} +
\frac{(3\Delta_{12} +3\Delta_{\beta_1} - 3\Delta_{\beta_2} + 1)^2}{3(1 + 3\Delta_{12})} 
\right\}A_\frac13
\ .
\end{align}
\label{csts}
\end{subequations}
We have computed $d_{\Delta_{12} + \frac k3}$ for $k\leq 6$, but we do not display the cases $k=5,6$ due to their lengthy expressions. Now, plugging the OPE~\eqref{OPE1} into~\eqref{sch}, we find the relation
\begin{equation}
d_{\Delta_{12} + \frac k3} = 
\hat{C}_{O_{\beta_1} O_{\beta_2} \mathcal{E}^{(k)}_{\beta_1 + \beta_2}}
\times R^{(k)}_{\beta_1 + \beta_2}
\quad\text{with}\quad
\langle \mathcal{E}^{(k)}_{\beta}(z) \rangle
= \frac{R^{(k)}_\beta}{|z-\bar z|^{2\Delta_{\beta} + \frac{2k}{3}}}
\ .
\label{one}
\end{equation}
In principle, we can compute the OPE coefficients $\hat{C}_{O_{\beta_1} O_{\beta_2} \mathcal{E}^{(k)}_{\beta_1 + \beta_2}}$ by considering the full-plane four-point function $\langle O_{\beta_1} O_{\beta_2} O_{\beta_1} O_{\beta_2} \rangle_{\mathbb{C}}$, and this was done for some special values of $\beta_1, \beta_2$ in~\cite{cfgk20}. For the generic case, the results of \cite{cfgk20} suggest that the structure constants in $\langle O_{\beta_1} O_{\beta_2} O_{\beta_1} O_{\beta_2} \rangle_{\mathbb{C}}$ may be written as a sum of products of the OPE coefficients, and we do not yet understand how this factorization works. For the special case $\langle O_\pi(z_1)O_{-\pi}(z_2)\rangle$, the situation is a bit better, the OPE coefficients $\hat{C}_{O_{\pi} O_{-\pi} \mathcal{E}^{(k)}_{0}}$ for $k\leq 6$ were determined in \cite{cfgk20} and take the form:
 \begin{align}
\hat C_{O_{\pi}O_{-\pi}
 \mathcal{E}^{(k)}_0} 
 = \frac{1}{k!} \left(\hat C_{O_{\pi}O_{-\pi}
 \mathcal{E}^{(1)}_0} 
\right)^{2k}
\quad\text{for} \quad k \leq6
\quad\text{with}\quad
\hat C_{O_{\pi}O_{-\pi}
 \mathcal{E}^{(1)}_0} 
 = \frac{3^{3/4} \sqrt{\frac{2 \pi }{5}} \sqrt{\lambda }}{\Gamma \left(-\frac{2}{3}\right)}
\ .
\label{coeff}
\end{align}
This allows us to compute $R_0^{(k)}$ in \eqref{one} for $k \leq 6$,
\begin{equation}
R_{0}^{(k)}  = 
(R_0^{(1)})^{k}
\quad\text{with}\quad
R_0^{(1)}
=
-e^{-\frac{4 \pi  \lambda }{5 \sqrt{3}}} \Gamma \left(-\frac{1}{3}\right)
\ .
\label{1pt}
\end{equation}
It is tempting to conjecture that \eqref{1pt} holds for any $k$, however, the pattern breaks down at $k=7$. More precisely, the OPE coefficient $\hat C_{O_{\pi}O_{-\pi} \mathcal{E}^{(7)}_0} $ is given in Eq.~(6.20) of \cite{cfgk20} and its expression is no longer of the type \eqref{coeff}. Using that expression, we have also computed $R_{0}^{(\frac73)}$, which turns out to be much more complicated than \eqref{1pt}.

Moreover, let us stress here that our computation suggests that $d_{\Delta_{12} + \frac k3}$ does not vanish for any $k$, and this implies that the operator $\mathcal{E}^{(k)}_\beta(z)$ has a non-vanishing one-point function, which is inconsistent with the mathematical definition of $\mathcal{E}^{(k)}_\beta(z)$ in \cite{cfgk21}. This means that the definition of  $\mathcal{E}^{(k)}_\beta(z)$ in~\cite{cfgk21} is not consistent with the OPE of the layering operators on the upper half-plane in~\eqref{OPE1}. In Section~\ref{sec:newE}, we will propose a new definition of $\mathcal{E}^{(k)}_\beta(z)$ which is consistent with the OPE \eqref{OPE1}.

\subsection{The $t$-channel expansion}
We proceed with the limit $z_2 \rightarrow \bar z_2$ of \eqref{2pts}, or equivalently, we send the cross-ratio $\sigma$ in~\eqref{cross} to 1. This limit will give us a new list of boundary operators in the Brownian loop soup. First, the limit $\sigma \rightarrow 1$ of $G_{\beta_1, \beta_2}(\sigma)$ implies that we can write
 \begin{equation} 
G_{\beta_1, \beta_2}(\sigma) \overset{\sigma \rightarrow 1}{=} (1 - \sigma)^{-2\Delta_{\beta_1} }
\left(
1 + 
\sum_{k =2}^\infty B_k (1 - \sigma)^{k}
\right)
\ ,
\label{G1}
\end{equation}
where the coefficients $B_k$ are always non-zero. Recall that the $s$-channel and $t$-channel conformal blocks are related by the relation 
\begin{equation}
\mathcal{F}_\Delta^{(t)}(\Delta_{\beta_1}, \Delta_{\beta_2}, \Delta_{\beta_3}, \Delta_{\beta_4} | \sigma)
=
\mathcal{F}_\Delta^{(s)}(\Delta_{\beta_1}, \Delta_{\beta_4}, \Delta_{\beta_3}, \Delta_{\beta_2} | 1- \sigma)
\ .
\label{rela}
\end{equation}
Therefore, using the above relation with the equations \eqref{sa} and \eqref{G1} implies that \eqref{G1} can be expanded into conformal blocks of non-negative integer dimensions. 
Thus, we assume that the $t$-channel decomposition of the two-point function \eqref{2pts} can be written as follows:
\begin{equation}
\langle 
O_{\beta_1}(z_1)
O_{\beta_2}(z_2)
\rangle_{\mathbb{H}}
\overset{z_2 \rightarrow \bar z_2}{=}
|z_1 - \bar z_2|^{-4\Delta_{\beta_1}}|z_2 - \bar z_2|^{2\Delta_{\beta_1} - 2\Delta_{\beta_2}}
 \sum_{k =0}^\infty D_k^{\beta_1, \beta_2} \mathcal{F}_k^{(t)}(\Delta_{\beta_1}, \Delta_{\beta_1}, \Delta_{\beta_2}, \Delta_{\beta_2}| \sigma)
\ .
\label{ant}
\end{equation}
Comparing \eqref{2pts}, \eqref{G1}, and \eqref{ant}, we obtain a linear system for the structure constants $D_k^{\beta_1, \beta_2}$. However, we refrain from writing it down explicitly since the computation is similar to the previous $s$-channel computation. For generic $\beta_1, \beta_2$, we find that $D_k^{\beta_1, \beta_2}$ does not vanish for any $k$. In terms of the OPE, this means that there are infinitely many boundary operators of non-negative integer dimension in the bulk-boundary OPE of $O_{\beta_1}(z_1)$, as written in \eqref{OPEbd}. Nevertheless, let us also display a few examples of $D_k^{\beta_1, \beta_2}$:
\begin{subequations}
\begin{align}
D_0^{\beta_1, \beta_2} &= 1
\ , \\
D_1^{\beta_1, \beta_2} &= 0
\ , \\
D_2^{\beta_1, \beta_2} &= \frac{1}{10} \left(\Delta _{\beta _1}+\Delta _{\beta _2}-\Delta
   _{12}\right)-\frac{\Delta _{\beta _1} \Delta _{\beta _2}}{\lambda}
\ , \\
D_3^{\beta_1, \beta_2} &=
 -\frac{1}{10} \left(\Delta _{\beta _1}+\Delta _{\beta _2}-\Delta
   _{12}\right)+\frac{ \Delta _{\beta _1} \Delta _{\beta _2}}{\lambda}
\ .
\end{align}
\end{subequations}
Notice that  $D_2^{\beta_1, \beta_2}$ and $D_3^{\beta_1, \beta_2}$ are zero for $\beta_2 = -\beta_1 = \pi $. For these special values of $\beta_1, \beta_2$, our results imply that
there are infinitely many vanishing structure constants $D_k^{\pi, -\pi}$. Up to $k \leq7$, we find
\begin{subequations}
\begin{align}
D_0^{\pi, -\pi} &= 1
\ , \\
D_1^{\pi, -\pi} &= D_2^{\pi, -\pi} = D_3^{\pi, -\pi} = D_5^{\pi, -\pi} = D_7^{\pi, -\pi} = 0 \label{D1-7}
\ , \\
D_4^{\pi, -\pi} &= \frac{18 \lambda ^2}{6875 (5 \lambda +11)}
\ , \\
D_6^{\pi, -\pi} &= \frac{\lambda ^2 (\lambda  (15599 \lambda -4362)-200)}{5843750 (\lambda +12) (4 \lambda -1) (7 \lambda
   +34)}
    \ .
\end{align}
\label{exD}
\end{subequations}
Therefore, we conjecture
\begin{equation}
D_k^{\pi, -\pi} = 0
\quad\text{for}\quad
k = 2
\;\;\text{and}\;\; 3,5,7, \ldots
\ .
\label{cond}
\end{equation}
Recalling that the two-point function in \eqref{2pts} is invariant under $\beta_i \rightarrow \beta_i + 2\pi n, n\in\mathbb{Z}$, Eq.~\eqref{cond} leads to the vanishing of the OPE coefficients in~\eqref{conb}.

\section{Probabilistic interpretation of $e^{(2k)}$ \label{sec:prob}}
In order to find an interpretation of the operators $e^{(2k)}$ in terms of Brownian loops, we rewrite the two-point function \eqref{2pts} in terms of the Brownian loop measure $\mu_{\mathbb{H}}^{\text{loop}}$, which appears in the partition function~\eqref{BLS} of the Brownian loop soup. 

From \cite{cgk16}, using the displayed equation just above Eq.~(4.3) of that paper, elementary calculations lead to
\begin{align} \label{eq:2-point-func-1}
    \langle {O}_{-\beta}(z_1){O}_{\beta}(z_2) \rangle_{\mathbb{H}} = \langle {O}_{-\beta}(z_1) \rangle_{\mathbb{H}} \langle {O}_{\beta}(z_2) \rangle_{\mathbb{H}} \, \exp{\big
    \{2\lambda(1-\cos\beta)\
    \mu_{\mathbb{H}}^{\text{loop}}(z_1,z_2)
    \big\}},
\end{align}
where 
$\langle {O}_{\beta}(z) \rangle = |z-\bar z|^{-2\Delta_\beta}$ and
$\mu_{\mathbb{H}}^{\text{loop}}(z_1,z_2)$
denotes the $\mu^{\text{loop}}_{\mathbb{H}}$-measure of all Brownian loops contained in the upper half-plane that wind around both $z_1$ and $z_2$, disconnecting them from infinity. There is an explicit expression for $\mu_{\mathbb{H}}^{\text{loop}}(z_1,z_2)$, which was derived rigorously in~\cite{hwz19}, but we refrain from displaying it to keep the discussion simple and concise. It is easy to check that, for $x_2\in\mathbb{R}$,
\begin{equation}
\mu_{\mathbb{H}}^{\text{loop}}(z_1,z_2)
\Big\vert_{z_2 = x_2 + i \varepsilon}
  = \frac{\varepsilon^{2}}{2} \, \mu^{\text{bub}}_{\mathbb{H}}(z_1, x_2) 
  +\ldots
     \label{eq:asymp}
  \ ,
\end{equation}
where $\mu^{\text{bub}}_{\mathbb{H}}$ is, up to a multiplicative constant, the so-called Brownian bubble measure introduced in \cite{lsw03}, and $\mu_{\mathbb{H}}^{\text{bub}}(z, x)$ is the $\mu_{\mathbb{H}}^{\text{bub}}$-measure of all Brownian excursions from $x$ to $x$ that disconnect $z$ from infinity. Furthermore, if $f$ is a conformal transformation of the upper half-plane to itself, we have that
\begin{equation}
\mu_{\mathbb{H}}^{\text{bub}}(z, x) = \vert f'(x) \vert^2 \mu_{\mathbb{H}}^{\text{bub}}(f(z), f(x))
\ .
\label{cont}
\end{equation}
That is to say, $\mu_{\mathbb{H}}^{\text{bub}}(z, x)$ transforms under conformal maps as a boundary quasi-primary operator of dimension 2. Combining \eqref{eq:2-point-func-1} and \eqref{eq:asymp}, we find
\begin{multline} 
\frac{
    \langle {O}_{-\beta}(z_1){O}_{\beta}(z_2) \rangle_{\mathbb{H}}}
    {
 \langle {O}_{\beta}(z_2) \rangle_{\mathbb{H}}}
     \Big\vert_{z_2 = x_2 + i \varepsilon}
= \langle {O}_{-\beta}(z_1) \rangle_{\mathbb{H}}
\Big\{
1 + \lambda(1-\cos\beta)
\mu_{\mathbb{H}}^{\text{bub}}(z_1, x_2)
\varepsilon^2
\\
 + \sum_{k=2}^{\infty}\frac{1}{k!}\lambda^k(1-\cos\beta)^k 
 \big(\mu_{\mathbb{H}}^{\text{bub}}(z_1, x_2) \big)^k \varepsilon^{2k} + \ldots
\Big\} \, .
\label{eq:2-point-func-2}
\end{multline}
Now let us insert the bulk-boundary OPE of $O_{\beta}(z_2)$, Eq.~\eqref{OPEbd}, inside the two-point function~\eqref{2pts}:
\begin{multline}
\frac{
\langle
O_{-\beta}(z_1)
O_{\beta}(z_2)
\rangle_{\mathbb{H}}
}{
 \langle {O}_{\beta}(z_2) \rangle_{\mathbb{H}}}
   \Big\vert_{z_2 = x_2 + i \varepsilon}
= \langle {O}_{-\beta}(z_1) \rangle_{\mathbb{H}}
 +
 \left\langle 
\left\{
 \frac{\Delta_{\beta}}{\lambda}
T(x_2)
+b^{(2)}_{\beta}Y(x_2)
\right\}
O_{-\beta}(z_1)
\right\rangle_{\mathbb{H}}
\left(2\varepsilon\right)^2
\\
+
\sum_{k =3}^\infty b^{(k)}_{\beta}
 \langle e^{(k)}(x_2)O_{\beta}(z_1)
 \rangle_{\mathbb{H}}
 (2\varepsilon)^{k}
 + \ldots \, ,
 \label{OPEe}
\end{multline}
where we have used $\varepsilon = \frac{z_2-\bar z_2}{2i}$. Next, recall the vanishing of the OPE coefficients $b^{(k)}_{\pi}$ for $k=2$ and $k$ odd, Eq.~\eqref{conb}. As a consequence, setting $\beta = \pi$ in \eqref{OPEe} leaves us with only the contribution from the boundary stress-energy tensor $T(x_2)$and the primary operators $e^{(2k)}(x_2)$, along with their descendants. Finally, let us compare the resulting OPE to \eqref{eq:2-point-func-2}.

For $\beta = \pi$, the boundary stress-energy tensor $T(x_2)$ is the only quasi-primary operator of dimension 2 on the right-hand side of \eqref{OPEe}. Using \eqref{cont}, we can then interpret $T(x_2)$ as an operator that inserts the outer boundary of a Brownian loop at $x_2$, as done in \cite{cfgk22}. In other words, we can write
\begin{align}
\mu_{\mathbb{H}}^{\text{bub}}(z_1, x_2)
\propto
\frac{\langle T(x_2) O_{-\pi}(z_1) \rangle_{\mathbb{H}}}{\langle  O_{-\pi}(z_1) \rangle_{\mathbb{H}}}
= \frac{\langle T(x_2) O_{\pi}(z_1) \rangle_{\mathbb{H}}}{\langle  O_{\pi}(z_1) \rangle_{\mathbb{H}}}
\ .
\label{Tmu}
\end{align}

Then, observe from \eqref{cont} that 
$\big(\mu_{\mathbb{H}}^{\text{bub}}(z_1, x_2) \big)^k$ transforms as a quasi-primary operator of dimension $2k$. For $\beta = \pi$, the only quasi-primary operator of dimension $2k$ on the right-hand side of \eqref{OPEe} is $e^{(2k)}(x_2)$~\footnote{Primary operators are quasi-primary operators, but the converse is not true.}. Therefore, we propose the physical interpretation of the primary operators $e^{(2k)}(x_2)$
as operators that insert the outer boundaries of $k$ independent Brownian loops at $x_2$. In other words, the bulk-boundary two-point functions $\langle e^{(2k)}(x_2) O_{\beta}(z_1) \rangle$ with $k\geq 3$ in \eqref{OPEe} correspond to powers of the $\mu_{\mathbb{H}}^{\text{bub}}$-measure in \eqref{eq:2-point-func-2}:
\begin{equation}
\left(\mu_{\mathbb{H}}^{\text{bub}}(z_1, x_2)
\right)^k
\propto
\frac{\langle e^{(2k)}(x_2) O_{\pi}(z_1) \rangle_{\mathbb{H}}}{\langle O_{\pi}(z_1) \rangle_{\mathbb{H}}}
\ .
\end{equation}
 





\section{Edge operators on the upper half-plane \label{sec:newE}}
In~\cite{cfgk21}, it is shown that the edge operator emerges from the OPE of two layering operators. This calculation is carried out on the full plane and the resulting edge operator has zero one-point function. The definition of the edge operator for more general domains given in~\cite{cfgk21} is inspired by the full-plane definition, so that the edge operator of~\cite{cfgk21} has zero one-point function in all domains. However, this is inconsistent with the upper half-plane OPE of the layering operators, Eq.~\eqref{OPE1}. For this reason, in this section we propose and analyze a new definition of the edge operator, which resolves the inconsistency.

\subsection{The edge operator of \cite{cfgk21}}
Given a Brownian loop soup in $D$, we denote by $\mathcal{N}_{\varepsilon}(z)$ the number of Brownian loops whose boundary comes $\varepsilon$-close to $z$.
Due to the scale invariance of the Brownian loop soup, $\mathcal{N}_{\varepsilon}(z)$ is actually infinite, but one can make sense of it by introducing an ultraviolet cutoff that removes the microscopically small loops. This procedure is carried out in~\cite{cfgk21}, where the edge operator $\mathcal{E}=\mathcal{E}_{CFGk}$ in a finite domain $D \subset \mathbb{C}$ is formally defined as
\begin{align}
    \mathcal{E}_{CFGK}(z) &=\lim_{\varepsilon\to 0} \vartheta_{\varepsilon}^{-1} \frac{\hat{c}}{\sqrt\lambda} \Big(\mathcal{N}_{\varepsilon}(z) - \langle \mathcal{N}_{\varepsilon}(z) \rangle_D \Big) 
    \nonumber \ , \\
    &= \lim_{\varepsilon\to 0} \vartheta_{\varepsilon}^{-1} \frac{\hat{c}}{\sqrt\lambda} \Big( \mathcal{N}(z) - \lambda\mu^{\text{loop}}_D(\ell \cap B_{\varepsilon}(z) \not=\emptyset) \Big),
    \label{defE}
\end{align}
where $\vartheta_{\varepsilon} \approx \varepsilon^{2/3}$ as $\varepsilon\to 0$, $B_{\varepsilon}(z)$ denotes the disk of radius $\varepsilon$ centered at $z$, and $\hat{c}$ is a constant chosen so that $\mathcal{E}_{CFGK}$ is canonically normalized on the full plane. Furthermore, we write $\mu^{\text{loop}}_D\big(\ell \cap B_{\varepsilon}(z) \not=\emptyset \big)$ for the $\mu^{\text{loop}}_D$-measure of Brownian loops that are contained in domain $D$ and whose boundaries intersect the disk $B_{\varepsilon}(z)$. Note that
\begin{equation}
    \langle \mathcal{N}_{\varepsilon}(z) \rangle_D = \lambda\mu^{\text{loop}}_D(\ell \cap B_{\varepsilon}(z) \not=\emptyset)
\end{equation}
follows immediately from the fact that the Brownian loop soup in $D$ is defined as a Poisson process with intensity measure $\lambda\mu^{\text{loop}}_D$.

From the definition~\eqref{defE}, it is clear that the one-point function of $\mathcal{E}_{CFGK}$ is zero for any domain $D$.
The higher-order edge operators $\mathcal{E}^{(k)}_0$ of~\cite{cfgk21}, with $k>1$, are essentially defined as the normal order of $(\mathcal{E}_{CFGK})^k$. As remarked in~\cite{cfgk21}, it follows immediately from the definition that the one-point function of $\mathcal{E}^{(k)}_0(z)$ is also zero in any domain.
The vanishing of these one-point functions is clearly incompatible with Eq.~\eqref{OPE1} for $\beta_2=-\beta_1$, so below we propose a new definition of the edge operators.
Note that the higher-order charged edge operator $\mathcal{E}^{(k)}_\beta$ is essentially the product of the higher-order edge operator $\mathcal{E}^{(k)}_0$ and the layering operator $O_{\beta}$, so its one-point function is not automatically zero. The one-point function of the simple charged edge operator in $D$, for example, is $\langle\mathcal{E}O_{\beta}\rangle_D$. Nevertheless, a redefinition of the uncharged edge operators would clearly also affect the charged ones.

\subsection{New definition of the edge operator}
We propose the following redefinition of the edge operator $\mathcal{E}(z)$ in $D$:
\begin{align} 
\label{def:new_E}
     \mathcal{E}(z)
    & = \mathcal{E}_{CFGK}(z) - \hat{c}\sqrt\lambda\lim_{\varepsilon\to 0}\vartheta_{\varepsilon}^{-1}\mu^{\text{loop}}_{\mathbb{C}}(\ell \cap B_{\varepsilon}(z) \not=\emptyset
    \;\;
    \text{and}
    \;\;
    \ell \cap (\mathbb{C}\setminus D) \not=\emptyset),
\end{align}
where $\mathcal{E}_{CFGK}$ denotes the edge operator as defined in~\cite{cfgk21} and $\mu^{\text{loop}}_{\mathbb{C}}(\ell \cap B_{\varepsilon}(z) \not=\emptyset
    \;\;
    \text{and}
    \;\;
    \ell \cap (\mathbb{C}\setminus D) \not=\emptyset)$ is the $\mu^{\text{loop}}_{\mathbb{C}}$-measure of Brownian loops in $\mathbb{C}$ whose boundaries $\ell$ intersect both $B_{\varepsilon}(z)$ and $\mathbb{C} \setminus D$.
The difference between the two definitions in a domain $D\neq\mathbb{C}$ is that in the new $\mathcal{E}(z)$ we subtract the weight of all loops in $\mathbb{C}$ coming close to $z$, not only of those contained in $D$.
With this definition, the one-point function of the edge operator in $D$ is
\begin{equation} \label{eq:1-point-func-edge-def}
    \langle \mathcal{E}(z) \rangle_D = -\hat{c}\sqrt{\lambda}\lim_{\varepsilon\to 0}\vartheta_{\varepsilon}^{-1}\mu^{\text{loop}}_{\mathbb{C}}(\ell \cap B_{\varepsilon}(z) \not=\emptyset
     \;\;
    \text{and}
    \;\;
    \ell \cap (\mathbb{C}\setminus D) \not=\emptyset).
\end{equation}

The proof of Lemma~2.2 of~\cite{cfgk21} implies that, for any point $z$ in the interior of $D$, the limit in~\eqref{eq:1-point-func-edge-def} exists and is conformally covariant in the sense that, for any $D'$ conformally equivalent to $D$,
\begin{equation} \label{eq:conf-cov}
    \langle \mathcal{E}(z') \rangle_{D'} = |f'(z)|^{-\frac23} \langle \mathcal{E}(z) \rangle_D,
\end{equation}
where $f$ is any conformal map from $D$ to $D'$ such that $f(z)=z'$.
It follows immediately from~\eqref{eq:conf-cov} that
\begin{equation} \label{eq:1-point-func-edge}
    \langle \mathcal{E}(z) \rangle_D = -\sqrt{\lambda} \, c_1 \, \text{rad}(z,D)^{-\frac23},
\end{equation}
where $c_1$ is a constant and $\text{rad}(z,D)$ denotes the conformal radius of the domain $D$ from point $z\in D$.
To see this, given $z\in D$, let $D'$ be the unit disk $\mathbb{D}$ and choose $f:D\to\mathbb{D}$ such that $f(z)=0$. Then~\eqref{eq:conf-cov} gives
\begin{align}
    \langle \mathcal{E}(z) \rangle_D &= \langle \mathcal{E}(0) \rangle_{\mathbb{D}} \, |f'(z)|^{\frac23}
\ , \nonumber\\    
    &= \langle \mathcal{E}(0) \rangle_{\mathbb{D}} \, \text{rad}(z,D)^{-\frac23}.
\end{align}
When $D$ is the upper half-plane $\mathbb{H}$, $\text{rad}(z,\mathbb{H})=|z-\bar z|$, so we have
\begin{equation} \label{eq:1-pt_fnc_E}
    \langle \mathcal{E}(z) \rangle_{\mathbb{H}} = \frac{\langle \mathcal{E}(0) \rangle_{\mathbb{D}}}{|z-\bar z|^{\frac23}} \, ,
\end{equation}
which is the expected behavior for the one-point function of a primary operator of scaling dimension $\frac23$. 

It is straightforward to modify the definition of the higher-order edge operators in the same way as the definition of $\mathcal{E}$, which results in operators whose one-point functions do not automatically vanish in all domains. This procedure removes the inconsistency between
the definition of the edge operators of~\cite{cfgk21} and
the OPE in~\eqref{OPE1}.


To conclude this section, we briefly consider the bulk-boundary OPE of the new edge operator. Deriving such an OPE typically requires calculating a two-point function containing $\mathcal{E}$. It is not immediate how to do this using CFT techniques, but using the definition of the edge operator in \eqref{def:new_E}, we can write $\mathcal{E}(z) = \mathcal{E}_{CFGK}(z) + \langle \mathcal{E}(z) \rangle$, which implies that
\begin{align}
    \langle \mathcal{E}(z)O_{\beta}(z') \rangle_{\mathbb{H}} &= \langle \mathcal{E}_{CFGK}(z) O_{\beta}(z') \rangle_{\mathbb{H}}
    + \langle\mathcal{E}(z)\rangle_{\mathbb{H}} \langle O_{\beta}(z')\rangle_{\mathbb{H}}
    \nonumber
    \\
    & = \langle \mathcal{E}_{CFGK}(z) O_{\beta}(z') \rangle_{\mathbb{H}} + R_{0}^{(1)} \, |z-\bar z|^{-\frac23} \langle O_{\beta}(z')\rangle_{\mathbb{H}} \, ,
\end{align}
where, in the second line, we have rewritten the one-point function $\langle\mathcal{E}(z)\rangle_{\mathbb{H}}$ in terms of $R_0^{(1)}$ as in~\eqref{one}.
Next, using loop-soup arguments similar to those in the proof of Theorem~3.1 of~\cite{cfgk21},} one can show that
\begin{align}
\begin{split}
    & \langle \mathcal{E}_{CFGK}(z) O_{\beta}(z') \rangle_{\mathbb{H}} \propto 
    \langle O_{\beta}(z') \rangle_{\mathbb{H}} \, \mu^{\text{bub}}_{\mathbb{H}}(z',x)|z- \bar z|^{\frac43}+ \ldots \, .
\end{split}
\end{align}
From the previous section, we know that the boundary stress-energy tensor $T (x)$ can be identified with an operator that inserts the outer boundary of a Brownian loop at point $x$, therefore using \eqref{Tmu} allows us to write
\begin{align}
     \langle \mathcal{E}(z)O_{\beta}(z') \rangle_{\mathbb{H}} &\overset{z\rightarrow \bar z}{=} 
       |z-\bar z|^{-\frac23} 
      \Big( R_0^{(1)}\langle O_{\beta}(z')\rangle_{\mathbb{H}} + C\langle T(x) O_{\beta}(z') \rangle_{\mathbb{H}} \, \vert z-\bar{z} \vert^2 + \ldots \Big) \, ,
\end{align}
where $C$ is a constant. From a CFT perspective, the constant $C$ is the bulk-boundary OPE coefficient between $T(x)$ and $\mathcal{E}(z)$. Using the CFT definition of the boundary stress-energy tensor $T(x)$, it is easy to show that $C=\frac{2\Delta^{(\mathcal{E})}}{c} = \frac{1}{3\lambda}$.
Therefore, the above expansion suggests the following bulk-boundary OPE of  $\mathcal{E}(z)$:
\begin{equation}
\mathcal{E}(z) \overset{z\rightarrow\bar z}{=} |z-\bar z|^{-\frac23}\Big(
R_{0}^{(1)}
+ 
\frac{2}{3\lambda}T(x) \, |z-\bar z|^2 + \ldots \Big) \, .
\label{newE}
\end{equation}
In contrast to the edge operator of \cite{cfgk21}, which behaves as $\mathcal{E}_{
\text{CFGK}
}(z)\overset{z\rightarrow \bar z}\sim T(x)|z -\bar z|^{\frac43}$, the leading operator of the OPE \eqref{newE} is the identity operator, due to the additional term in \eqref{def:new_E}.



\section{Outlook \label{sec:out}}

In this note, we obtain a set of new boundary operators, $Y(x)$ and $e^{(k)}(x)$ in \eqref{loc}, for the Brownian loop soup on the upper half-plane. We show that $e^{(2k)}(x)$ can be interpreted as an operator that inserts $k$ outer boundaries of Brownian loops at the point $x$ on the real axis. The physical interpretation of $Y(x)$ and $e^{(2k - 1)}(x)$ remains unclear.

\subsubsection{Correlation functions of $e^{(2k)}(x)$}
Since we already understand the physical interpretation of $e^{(2k)}(x)$, it would be interesting to compute their correlation functions, which, in principle, can be expressed in terms of the $\mu^{\text{bub}}_{\mathbb{H}}$-measure, as was done for $\langle e^{(2k)}(x_1) O_{\beta}(z_2) \rangle$ in Section \ref{sec:prob}. For the bulk counterparts $\mathcal{E}_{0}^{(k)}(z)$, which insert the outer boundaries of $k$ Brownian loops at the point $z$, the authors of \cite{cfgk21} have written down explicitly the correlation functions of $\mathcal{E}_{0}^{(k)}(z)$ in terms of the Brownian loop measure $\mu_{\mathbb{H}}$ that appears in \eqref{BLS}. We leave the issue of finding the relation between correlation functions of $e^{(2k)}(x)$ and $\mu^{\text{bub}}_{\mathbb{H}}$, as well as their expressions, for future work.

\subsubsection{Towards a CFT description of the Brownian loop soup}
While we have made progress in listing the boundary local operators in the spectrum of the CFT associated to the Brownian loop soup, understanding the CFT spectrum amounts to knowing not only the list of local operators, but also the model's symmetries and the domains of all physical parameters. The last two points have not yet been fully understood. Here, we discuss possible ways of moving beyond the current knowledge.

\begin{itemize}
    \item \textbf{Physical domain of $\beta$:}
    Any function satisfying the conformal Ward identities can be decomposed into Virasoro conformal blocks. Nevertheless, the ability to decompose a collection of functions into conformal blocks does not automatically guarantee that those functions lead to consistent OPEs and are the $n$-point functions of a full-fledged CFT. This observation is relevant for the layering operator. In~\cite{cgk16}, where the layering operator was introduced, it was shown that the $n$-point functions $\langle O_{\beta_1}(z_1) \ldots O_{\beta_n}(z_n) \rangle$ exist and are conformally covariant for all $n$ and all choices of parameters $\beta_1,\ldots,\beta_n$. However, on the full plane, those $n$-point functions are nonzero only if $\sum_i\beta_i=2\pi n, n\in\mathbb{Z}$, a condition reminiscent of the charge neutrality condition that guarantees that the $n$-point functions of vertex operators do not vanish.
Moreover, from previous results of~\cite{cfgk20}, we observe that some structure constants of correlation functions of $ O_{\beta}(z)$ for $\beta=\pi n, n\in\mathbb{{Z}}$, admit simple factorizations in terms of the OPE coefficients. The same also happens for the upper half-plane two-point function $\langle O_{\pi}(z_1)O_{-\pi}(z_2) \rangle$, in which case some structure constants factorize into a product of OPE coefficients from \cite{cfgk20} and the coefficients in~\eqref{1pt} from the one-point functions in~\eqref{one}. This may suggest that it is possible to construct a consistent CFT describing various observables of the Brownian loop soup whose spectrum  contains only layering operators $O_\beta(z)$ with $\beta = \pi n, n\in\mathbb{Z}$. We suspect that the layering operators corresponding to other values of $\beta$ may not be part of the spectrum of a consistent CFT.

    \item \textbf{The model's symmetries:}
The authors of \cite{cfgk20} have shown that the spectrum of the putative CFT describing the Brownian loop soup contains infinitely many bulk primary operators, whose scaling dimensions are positive integers. From Noether's theorem, the existence of a primary operator of scaling dimension 1 is associated to a continuous global symmetry. In the Brownian loop soup, it remains unclear what the corresponding global symmetry is. Furthermore, the existence of primary operators of higher integer dimensions could imply higher symmetries. For instance, the 3-state Potts model has a primary operator of scaling dimension 3 associated to the $\mathcal{W}_3$--symmetry algebra, the $\mathfrak{sl}_3$--generalization of the Virasoro algebra \cite{fms97}. It also remains unclear whether the Brownian loop soup has any higher symmetry associated to the primary operators of higher integer dimensions.
Perhaps, the first step to understand the model's symmetries would be to study the correlation functions of the primary operator of dimension 1. In particular, based on~\cite{cz02, car93}, we expect that the limit $\lambda \rightarrow 0$ of the two-point function of that operator is related to the area density of the region surrounded by the outer boundary of a Brownian loop.

\end{itemize}

\section*{Acknowledgments}
This project was started in the summer of 2025 during a visit to the Hausdorff Research Institute for Mathematics (HIM) in Bonn.
We are grateful to HIM for its hospitality during the Trimester Program ``Probabilistic methods in quantum field theory'' funded by the Deutsche Forschungsgemeinschaft (DFG, German Research Foundation) under Germany's Excellence Strategy - EXC-2047/1 - 390685813.
R.~N. thanks Max Downing, Jesper Lykke Jacobsen, Sylvain Ribault, Hubert Saleur for collaborating on related projects, and Xin Sun and Gefei Cai for discussions on the Brownian loop soup.
\appendix
\section{Conformal blocks \label{ap:Vir}}
We explain how to compute the coefficients $a_k$ of the conformal block $\mathcal{F}_\Delta^{(s)}(\Delta_{\beta_1}, \Delta_{\beta_2}, \Delta_{\beta_3}, \Delta_{\beta_4} | \sigma)$ in \eqref{sa}.  In principle, determining $a_k$ requires computing all the contributions from the level-$k$ descendants of the chiral primary field $o_\Delta$ to that conformal block, and the computation gets tedious very quickly since the number of descendants at level $k$ equals to the size of integer partitions of $k$, which grows very fast.

One way to overcome this technical difficulty is to use the Zamolodchikov recursion of \cite{zam84}. In this recursion, we consider the conformal block as an infinite power series of the nome function $q(\sigma)$. For $\sigma\in\mathbb{C}\setminus\{0,1\}$, we have
\begin{equation}
q(\sigma) = \text{exp}\left\{
-\pi \frac{{}_2F_1\left(\frac12,\frac12;1;1-\sigma\right)}{{}_2F_1\left(\frac12,\frac12;1;\sigma\right)}
\right\}
\ ,
\end{equation}
In this appendix only, we parametrize the central charge $c$ and the conformal dimensions $\Delta$ as follows:
\begin{equation}
\Delta(P) = \frac{c-1}{24} + P^2
\quad\text{and}\quad
c = 1 - 6\left(b - b^{-1} \right)^2
\, .
\end{equation}
 The expression for the $s$-channel conformal block is then 
\begin{multline}
\mathcal{F}_\Delta^{(s)}(\Delta_{\beta_1}, \Delta_{\beta_2}, \Delta_{\beta_3}, \Delta_{\beta_4} | \sigma)
=
(16q)^{\Delta -\frac{c-1}{24}}
\sigma^{\frac{c-1}{24} - \Delta_{\beta_1} - \Delta_{\beta_2}}
(1 - \sigma)^{\frac{c-1}{24} - \Delta_{\beta_1} - \Delta_{\beta_4}}
\\
\times
\theta_3(q)^{ - 4(\Delta_{\beta_1} + \Delta_{\beta_2} + \Delta_{\beta_3} + \Delta_{\beta_4})}
\mathcal{H}_\Delta^{(s)}(\Delta_{\beta_1}, \Delta_{\beta_2}, \Delta_{\beta_3}, \Delta_{\beta_4}| \sigma)
\ ,
\label{Fsig}
\end{multline}
where $\theta_3(q)$ is the Jacobi theta function given by $\theta_3(q) = \displaystyle\sum_{n = -\infty}^\infty q^{n^2}$. Furthermore, the function $\mathcal{H}_\Delta^{(s)}(\Delta_{\beta_1}, \Delta_{\beta_2}, \Delta_{\beta_3}, \Delta_{\beta_4} | \sigma)$ is defined through the recursion
\begin{equation}
\mathcal{H}_\Delta^{(s)}(\Delta_{\beta_1}, \Delta_{\beta_2}, \Delta_{\beta_3}, \Delta_{\beta_4} | \sigma)
= 1 
+
 \sum_{m = 1}^\infty
 \sum_{n = 1}^\infty
\frac{(16q)^{mn}}{\Delta - \Delta_{(m,n)}}R_{m,n}
\mathcal{H}_{\Delta_{(m,-n)}}^{(s)}(\Delta_{\beta_1}, \Delta_{\beta_2}, \Delta_{\beta_3}, \Delta_{\beta_4} | \sigma)
\ .
\end{equation}
With $\Delta_{(r,s)} = \Delta(P_{(r,s)})$ and $P_{(r,s)} = \frac12(rb - sb^{-1})$, the coefficient $R_{m,n}
$ is given by
\begin{multline}
R_{m,n}
= \frac{-2P_{(0,0)}P_{(m,n)}}{\prod_{r= 1-m}^m\prod_{s=1-n}^n 2P_{(r,s)}}\prod_{\pm}
\prod_{r\overset2=1-m}^{m-1}\prod_{r\overset2=1-m}^{m-1}
(P_2 \pm P_1 + P_{(r,s)})(P_3 \pm P_4 +P_{(r,s)})
\ .
\label{R}
\end{multline}
The appearance of $P_{(0,0)}$ above is an abuse of notation to stress that the product in the denominator of \eqref{R} skips the term with $(m,n) = (0,0)$. The expression \eqref{Fsig} is  efficient for computing the conformal blocks as a power series in $\sigma$. Let us display the leading terms:
\begin{equation}
\mathcal{F}_\Delta^{(s)}(\Delta_{\beta_1}, \Delta_{\beta_2}, \Delta_{\beta_3}, \Delta_{\beta_4} | \sigma)
\overset{\sigma \rightarrow 0}= \sigma^{\Delta - \Delta_{\beta_1} - \Delta_{\beta_2}}
\left(
1 + \frac{(\Delta_{\beta_1} - \Delta_{\beta_2} + \Delta)(\Delta + \Delta_{\beta_4} - \Delta_{\beta_3})}{2\Delta}\sigma + \ldots
\right)
\ .
\end{equation}
With \eqref{Fsig}, we can easily obtain the analytic expression for the coefficients of the terms up to order $\sigma^8$ by using brute force computation.

\bibliographystyle{morder8}
\bibliography{992}

@article{LW03,
    author = {Lawler, Gregory F. and Werner, Wendelin},
    title = {The Brownian loop soup},
    journal = {Probability Theory and Related Fields},
    year = {2004},
    volume = {128},
    number = {4},
    pages = {565--588},
    doi = {10.1007/s00440-003-0319-6},
    url ={https://doi.org/10.1007/s00440-003-0319-6}
}

@article{lsw03,
   title={Conformal restriction: The chordal case},
   volume={16},
   ISSN={1088-6834},
   url={http://dx.doi.org/10.1090/S0894-0347-03-00430-2},
   DOI={10.1090/s0894-0347-03-00430-2},
   number={4},
   journal={Journal of the American Mathematical Society},
   publisher={American Mathematical Society (AMS)},
   author={Lawler, Gregory and Schramm, Oded and Werner, Wendelin},
   year={2003},
   month=jun, pages={917–955} }

@article{w06,
      title={The conformally invariant measure on self-avoiding loops}, 
      volume = {21},
      pages= {137-168},
      journal={Journal of the American Mathematical Society},
      publisher={American Mathematical Society (AMS)},
      author={Wendelin Werner},
      year={2008},
      url={https://arxiv.org/abs/math/0511605}, 
      DOI={10.1090/S0894-0347-07-00557-7}

}

@article{hwz19,
      title={On The Brownian Loop Measure},
      journal = {Journal of Statistical Physics},
      volume = {175},
      pages = {987--1005},
      author={Yong Han and Yuefei Wang and Michel Zinsmeister},
      year={2017},
      url={https://arxiv.org/abs/1707.00965}, 
      DOI = {10.1007/s10955-019-02275-7}
}

@article{w03,
      title = {SLEs as boundaries of clusters of Brownian loops},
      journal = {Comptes Rendus Mathematique},
      volume = {337},
      number = {7},
      pages = {481-486},
      year = {2003},
      issn = {1631-073X},
      doi = {10.1016/j.crma.2003.08.003},
      author={Wendelin Werner},
      url={https://arxiv.org/abs/math/0308164}, 
}

@article{bc10,
      title = {Universal Behavior of Connectivity Properties in Fractal Percolation Models},
      volume = {15},
      journal = {Electronic Journal of Probability},
      publisher = {Institute of Mathematical Statistics and Bernoulli Society},
      pages = {1394 -- 1414},
      year = {2010},
      doi = {10.1214/EJP.v15-805},
      URL = {https://doi.org/10.1214/EJP.v15-805},
      author={Erik I. Broman and Federico Camia},
      }

@article{cs09,
   title={Twist operator correlation functions in O(n) loop models},
   volume={42},
   ISSN={1751-8121},
   url={http://dx.doi.org/10.1088/1751-8113/42/23/235001},
   DOI={10.1088/1751-8113/42/23/235001},
   number={23},
   journal={Journal of Physics A: Mathematical and Theoretical},
   publisher={IOP Publishing},
   author={Simmons, Jacob J H and Cardy, John},
   year={2009},
   month=may, pages={235001} }

@article{cardy84,
title = {Conformal invariance and surface critical behavior},
journal = {Nuclear Physics B},
volume = {240},
number = {4},
pages = {514-532},
year = {1984},
issn = {0550-3213},
doi = {10.1016/0550-3213(84)90241-4},
url = {https://www.sciencedirect.com/science/article/pii/0550321384902414},
author = {John L. Cardy}
}

@article{cn04,
    title={Continuum Nonsimple Loops and 2D Critical Percolation},
    volume={116},
    DOI={10.1023/B:JOSS.0000037221.31328.75},
    number={},
    journal={Journal of Statistical Physics},
    pages={157--173},
    publisher={Springer Nature},
    author={Camia, Federico and Newman, Charles M.},
    year={2004},
    month=aug }

@article{cn07,
    title={Critical percolation exploration path and SLE$_6$: a proof of convergence},
    volume={139},
    ISSN={473–519},
    DOI={10.1007/s00440-006-0049-7},
    number={},
    journal={Probab. Theory Relat. Fields},
    pages={473--519},
    publisher={Springer Nature},
    author={Camia, Federico and Newman, Charles M.},
    year={2004},
    month=aug }

@article{cfgk22,
   title={The Brownian loop soup stress-energy tensor},
   volume={2022},
   ISSN={1029-8479},
   DOI={10.1007/jhep11(2022)009},
   number={11},
   journal={Journal of High Energy Physics},
   publisher={Springer Science and Business Media LLC},
   author={Camia, Federico and Foit, Valentino F. and Gandolfi, Alberto and Kleban, Matthew},
   year={2022},
   month=nov }

@article{cfgk21,
   title={Scalar Conformal Primary Fields in the Brownian Loop Soup},
   volume={400},
   ISSN={1432-0916},
   url={http://dx.doi.org/10.1007/s00220-022-04611-7},
   DOI={10.1007/s00220-022-04611-7},
   number={2},
   journal={Communications in Mathematical Physics},
   publisher={Springer Science and Business Media LLC},
   author={Camia, Federico and Foit, Valentino F. and Gandolfi, Alberto and Kleban, Matthew},
   year={2022},
   month=dec, pages={977--1018} }

@article{cgk16,
   title={Conformal correlation functions in the Brownian loop soup},
   volume={902},
   ISSN={0550-3213},
   url={http://dx.doi.org/10.1016/j.nuclphysb.2015.11.022},
   DOI={10.1016/j.nuclphysb.2015.11.022},
   journal={Nuclear Physics B},
   publisher={Elsevier BV},
   author={Camia, Federico and Gandolfi, Alberto and Kleban, Matthew},
   year={2016},
   month=jan, pages={483--507} }

@article{cfgk20,
   title={Exact correlation functions in the Brownian Loop Soup},
   volume={2020},
   ISSN={1029-8479},
   url={http://dx.doi.org/10.1007/JHEP07(2020)067},
   DOI={10.1007/jhep07(2020)067},
   number={67},
   journal={Journal of High Energy Physics},
   publisher={Springer Science and Business Media LLC},
   author={Camia, Federico and Foit, Valentino F. and Gandolfi, Alberto and Kleban, Matthew},
   year={2020},
   month=jul }

@article{schram99,
      title={Scaling limits of loop-erased random walks and uniform spanning trees}, 
      author={Oded Schramm},
      journal={Israel Journal of Mathematics},
      volume={118},
      pages={221-288},
      year={2000},
      DOI={10.1007/BF02803524}
}

@article{smirnov01,
    author = {Stanislav Smirnov}, 
    title = {Critical percolation in the plane: conformal invariance, Cardy's formula, scaling limits},
    journal = {Comptes Rendus de l'Académie des Sciences - Series I - Mathematics},
    volume = {333},
    pages = {239-244},
    year = {2001},
    DOI = {10.1016/S0764-4442(01)01991-7}
}

@article{sw11,
      title={Conformal Loop Ensembles: The Markovian characterization and the loop-soup construction},
      journal = {Annals of Mathematics},
      volume = {176},
      pages = {1827-1917},
      author={Scott Sheffield and Wendelin Werner},
      year={2012},
      url={https://arxiv.org/abs/1006.2374},
      DOI = {10.4007/annals.2012.176.3.8},
}

@article{CDCHKS14,
    author = {Dmitry Chelkak and Hugo Duminil-Copin and Clément Hongler and Antti Kemppainen and Stanislav Smirnov},
    title = {Convergence of Ising interfaces to Schramms SLE curves},
    journal = {Comptes Rendus. Mathematique},
    volume = {352},
    number = {2},
    pages = {157-161},
    year = {2014},
    DOI = {10.1016/j.crma.2013.12.002}
}

@article{she06,
      title={Exploration trees and conformal loop ensembles}, 
      author={Scott Sheffield},
      journal={Duke Math. J.},
      volume={147},
      number={1},
      pages={79-129},
      DOI={10.1215/00127094-2009-007},
      year={2009},
}

@article{car93,
	archiveprefix = {arXiv},
	author = {Cardy, John L.},
	doi = {10.1103/PhysRevLett.72.1580},
	journal = {Phys. Rev. Lett.},
	pages = {1580--1583},
	reportnumber = {OUTP-93-34S, OUTP-93-??S},
	title = {{Mean area of selfavoiding loops}},
	volume = {72},
	year = {1994}}

@article{cz02,
	author = {Cardy, John and Ziff, Robert},
	publisher = {arXiv},
	title = {Exact results for the universal area distribution of clusters in percolation, Ising and Potts models},
    journal = {Journal of Statistical Physics},
    volume = {110},
    year = {2003},
    pages = {1--33},
    DOI = {10.1023/A:1021069209656},
	url = {https://arxiv.org/abs/cond-mat/0205404},
	year = {2002}}

@article{Nienhuis82,
	author = {Nienhuis, Bernard},
	doi = {10.1103/PhysRevLett.49.1062},
	issue = {15},
	journal = {Phys. Rev. Lett.},
	month = {Oct},
	numpages = {0},
	pages = {1062--1065},
	publisher = {American Physical Society},
	title = {Exact critical point and critical exponents of $\mathrm{O}(n)$ models in two dimensions},
	url = {https://link.aps.org/doi/10.1103/PhysRevLett.49.1062},
	volume = {49},
	year = {1982}}

@article{rib14,
	archiveprefix = {arXiv},
	author = {Ribault, Sylvain},
	eprint = {1406.4290},
	primaryclass = {hep-th},
	slaccitation = {%%CITATION = ARXIV:1406.4290;%%},
	title = {{Conformal field theory on the plane}},
doi = {10.48550/arXiv.1406.4290},
	type = {review},
	year = {2014}}

@article{zam84,
	author = {Zamolodchikov, Al.B.},
	doi = {10.1007/BF01214585},
	journal = {Commun. Math. Phys.},
	pages = {419-422},
	title = {{Conformal symmetry in two dimensions: an explicit recurrence formula for the conformal partial wave amplitude}},
	volume = {96},
	year = {1984}}

@book{fms97,
	author = {Di Francesco, P. and Mathieu, P. and S{\'e}n{\'e}chal, D.},
	doi = {10.1007/978-1-4612-2256-9},
	title = {Conformal field theory},
	year = {1997}}

@article{bpz84,
	author = {Belavin, A. A. and Polyakov, Alexander M. and Zamolodchikov, A. B.},
	doi = {10.1016/0550-3213(84)90052-X},
	journal = {Nucl. Phys.},
	pages = {333-380},
	reportnumber = {CERN-TH-3827},
	title = {{Infinite conformal symmetry in two-dimensional quantum field theory}},
	volume = {B241},
	year = {1984}}

\end{document}